\documentclass[aps,prl,twocolumn,showpacs]{revtex4}
\usepackage{amssymb}

\usepackage{graphicx}

\begin{document}

\title{Signatures of Long-Range Spin-Triplet Component in Andreev
Interferometer}
\author{Anatoly F. Volkov}
\affiliation{Institut f\"ur Theoretische Physik III, Ruhr-Universit\"at Bochum, D-44801
Bochum, Germany}
\date{\today }

\begin{abstract}
We analyze the Josephson,$I_{J}$, and dissipative,$I_{V}$, currents in a
magnetic Andreev interferometer in the presence of the long-range spin
triplet component (LRSTC). Andreev interferometer has a cross-like geometry
and consists of a SF$_{l}$ - F - F$_{r}$S circuit and perpendicular to it a
N - F - N circuit, where S, F$_{l,r}$ are superconductors and weak
ferromagnets with non-collinear magnetisations $\mathbf{M}_{l,r}$, F is a
strong ferromagnet. The ferromagnetic wire F can be replaced with a
non-magnetic wire n. In the limit of a weak proximity effect (PE), we obtain
simple analytical expressions for the currents $I_{J}$ and $I_{V}$. In
particular, the critical Josephson current in a long Josephson junction (JJ)
is $I_{c}(\alpha ,\beta )=I_{0c}\chi (\alpha ,\beta )$, where the function $%
\chi (\alpha ,\beta )$ is a function of angles $(\alpha ,\beta )_{l,r}$ that
characterize the orientations of $\mathbf{M}_{l,r}$. The oscillating part of
the dissipative current $I_{osc}(V)=\chi (\alpha ,\beta )\cos \varphi
I_{0}(V)$ in the N - F(n)- N circuit depends on the angles $(\alpha ,\beta
)_{l,r}$ in the same way as the critical Josephson current $I_{c}(\alpha
,\beta )$, but can be much greater than the $I_{c}(\alpha ,\beta )$. At some
angles the current $I_{c}(\alpha ,\beta )$ changes sign. We briefly discuss
a relation between the negative current $I_{c}$ and paramagnetic response.
We argue that the measurements of the conductance in N - F(n) - N circuit
can be used as another complementary method to identify the LRSTC in S/F
heterostructures.
\end{abstract}

\maketitle
\affiliation{Theoretische Physik III,\\
Ruhr-Universit\"{a}t Bochum, D-44780 Bochum, Germany}
\date{\today }

\section{Introduction}

The phenomenon of phase coherence in superconducting systems is especially
well studied in Josephson junctions (JJs). In particular, if the magnetic
flux $\Phi $ of an external magnetic field $H_{ex}$ in a JJ with planar
geometry is equal to an integer number of the flux quanta $\Phi _{0}=hc/2e$ (%
$\Phi =n\Phi _{0}$), the Josephson current $I_{J}$ turns periodically to
zero \cite{KulikBook70,LikharevRMP79,BaroneBook82}. Another example of phase
coherence are the so-called Shapiro steps that arise on the $I-V$
characteristics in a JJ irradiated by an $ac$ electromagnetic field with a
frequency $\omega $. The positions of the steps $V_{n}$ is defined by the
condition $V_{n}=n\hbar \omega /2e$. Since the discovery of Josephson effect %
\cite{Josephson62}, various aspects of this effect have been intensively
studied on JJs of different types such as SIS, SNS, ScS junctions, where S,
N, c stand for a superconductor, normal metal and constriction, respectively%
\cite{KulikBook70,LikharevRMP79,BaroneBook82}.

In the last few decades a great attention was paid to the study of magnetic
JJs, \textit{i.e.} SFS junctions, where the Josephson coupling is realized
via a ferromagnetic layer(s) F. A number of interesting phenomena have been
predicted and observed in such JJs. One of them is the sign-reversal of the
Josephson critical current with changing temperature or thickness of the F
layer \cite{GolubovRMP04,BuzdinRev05,BVErev05,EschrigRev11,LinderRev15}.
This effect was originally predicted back in the 80s of the 20th century %
\cite{Bulaev77,Bulaev82,BuzdinKup90,BuzdinKup91}, but was observed
experimentally only much later \cite%
{RyazanovPRL01,RyazanovPRL06,ApriliPRL02,SellierPRB03,Weides06}.

Another interesting feature of magnetic JJs is the appearance of the
so-called long-range spin triplet component (LRSTC) of the condensate \cite%
{BuzdinRev05,BVErev05,EschrigRev11,LinderRev15,LinderBalRMP17}. The triplet
component is induced by proximity effect in the F layer in any magnetic JJs
due to Zeeman interaction of quasiparticles, which build Cooper pairs, with
an exchange field of a ferromagnet. However, a uniform exchange field
produces only a short-ranged component, which quickly decays inside the F
layer. The wave function of this component, $f$, is given by $f(t,t^{\prime
})$ $\sim \langle \psi _{\uparrow }(t)\psi _{\downarrow }(t^{\prime })+\psi
_{\downarrow }(t^{\prime })\psi _{\uparrow }(t)\rangle $ and its spin is
perpendicular to the magnetization vector $\mathbf{M}$ in F$.$ Such pairs
penetrate ferromagnet on a short distance of the order $\xi _{F}\cong \sqrt{%
D_{F}/E_{F}}$, where $D_{F}$ is the diffusion coefficient in F and $E_{F}$
is the exchange energy. In addition, this penetration is accompanied by
oscillations of $f(x)$ in space. At the same time, the actual LRSTC
described by the wave functions $f\sim \langle \psi _{\uparrow }(t)\psi
_{\uparrow }(t^{\prime })\rangle $ or $f\sim \langle \psi _{\downarrow
}(t)\psi _{\downarrow }(t^{\prime })\rangle $ occurs in magnetic JJs with a
non-uniform magnetization $\mathbf{M}(\mathbf{r})$ in the F film \cite%
{BVEprl01,Footnote,BuzdinRev05,BVErev05,EschrigRev11,LinderRev15,LinderBalRMP17}%
. The penetration depth of the LRSTC into a ferromagnet is much longer than $%
\xi _{F}$ and may be of the order of the Cooper pair penetration length into
a normal metal, $\xi _{T}=\sqrt{D_{F}/2\pi T}$. The prediction of a
long-range penetration of the triplet Cooper pairs into a ferromagnet was
observed in multiple experiments \cite%
{KlapwijkNature06,SosninVolkovPRL06,BirgePRL10,BlamireScience10,ZabelPRB10,AartsPRB10,BlamirePRL10, PetrashovZaikinPRL11,BirgePRL16,BirgePRB18,BlamirePRB18}%
. Observe that both components, long- and short-range, can be described by
the Fourier transform, $f_{\omega }$, of the function $f(t,t^{\prime })$
which should be an odd function of $\omega $ to satisfy the Pauli principle,
\textit{i.e.} $f(t,t)=0$ \cite{BuzdinRev05,BVErev05,EschrigRev11,LinderRev15}%
. The sign reversal of the critical Josephson current $I_{c}$ may be then
related both to the short-range (see reviews \cite%
{GolubovRMP04,BuzdinRev05,LinderRev15}) and long-range triplet components %
\cite%
{BVEprl03,AnischPRB06,EschrigPRB07,EschrigPRL08,Kawabata10,GolubovPRL10,HaltermanPRB14,HaltermanPRB15,BlamirePRL19}%
. The interest in the study of magnetic JJs is caused not only by new
physical effects, but also by possible applications of these junctions in
spintronics (see review \cite{LinderRev15} as well as recent papers \cite%
{Birge20,Blamire20,BlamireBuzdin20} and references therein) or in Josephson
magnetic random access memory \cite{Birge19}.

Although it is less known outside of the community, the phase coherence
takes place not only in JJs, but also in multi-terminal superconducting
structures like the so-called Andreev interferometers (see Fig.1), which for
a number of applications may have several important advantages over some
devices based on JJs \cite%
{Petrashov94,KlapwijkPRL96,TakayanVolkovPRL96,PannetierPRL96,PannVolkovPRL96,PetrashovPRL99,PetrashovPRB00, PetrashovPRL05,ErvandPRB10,ZaikinPRB12,GiazottoPRB17,ZaikinPRB19}%
. It has been found that the conductance between the N-reservoirs oscillates
with variation of the phase difference $\varphi $ between the
superconducting reservoirs S. The phase variation is provided either by
passing a $dc$ current between S reservoirs or by an external magnetic field
$H_{ext}$ applied in a superconducting loop connecting the S reservoirs.
Beside the conductance oscillations other interesting phenomena may arise in
Andreev interferometers \cite%
{TakayanVolkovPRL96,KlapwijkPRL96,PetrashovPRB00,
PetrashovPRL05,ZaikinPRB12,GiazottoPRB17,ZaikinPRB19} like the change of
sign of the Josephson critical current $I_{c}$ in multiterminal S-n-N
structures. In contrast to the change of sign discussed above for the
magnetic JJs, here it is related to an imbalance between the condensate and
the quasiparticles in the S-n-S circuit out-of-equilibrium. The effect of
sign inversion in S-n-S JJs has been considered in Refs. \cite{BagwellPRB97}
in ballistic JJs (see also references in \cite{ShumeikoPRB03}). In the more
practical case of diffusive JJs the sign change effect has been predicted in %
\cite{VolkovPRL95} (for further development of this idea, see Refs.\cite%
{ZaikinPRL98,Yip98, Bobkov10,Bobkov12}). The predicted effect has been
observed by the Klapwijk group \cite{vanWees02,KlapwijkNat06}. The voltage $%
V $ applied between N and S reservoirs leads to a non-equilibrium
distribution function $n(V)$ which affects strongly the current $I_{c}$ if $%
V\approx \Delta /e$.

Despite of numerous studies of various phenomena in magnetic superconducting
heterostructures, the LRSTC in Andreev interferometer remains largely
unexplored except for the conductance analysis in an interferometer-like
(three-terminal) superconducting system with a topological insulator and in
presence of a spin-orbit and Zeeman interactions \cite{Mal'shukovPRB18}. In
this manuscript we study the propagation of the LRSTC in an Andreev
interferometer and its influence on the conductance $G_{NN}$ of the N - F(n)
- N circuit as well as on the Josephson current $I_{J}$. As we are mostly
interested in propagation of the LRSTC, our results are equally applied for
the normal (n) or ferromagnetic (F) wire between the reservoirs S or N. It
is only assumed that its length, $L_{x}$, is larger than $\xi _{F}$ such
that only the LRSTC penetrates the F wire. Note also, that in the case of SF$%
_{l}$ - n - F$_{r}$S junction (horizontal line), not only the LRSTC
penetrates in the n-wire but also a spin singlet component.

The calculations are carried out in the approximation of a weak PE. We show
that the conductance $G_{NN}$ contains a part $G_{osc}$ which oscillates
with the increase of the phase difference $\varphi $: $G_{osc}=G_{0}\chi
(\alpha ,\beta )\cos \varphi $, where $\chi (\alpha ,\beta )$ is a function
of the angles $(\alpha ,\beta )$ which characterize the magnetization
vectors in the ferromagnetic layers $F_{l,r}$.(see Eq.(\ref{J3Ang})). The
Josephson current $I_{J}=I_{c}(\alpha ,\beta )\sin \varphi $ has the
standard phase dependence with the critical current $I_{c}(\alpha ,\beta )$
which has the same angle dependence as $G_{osc}(\alpha ,\beta )$. In the
case of SF$_{l}$ - F - F$_{r}$ structure, the critical current turns to zero
at $\alpha _{l}=\alpha _{r}\pm \pi /2$ and any $\beta $ or at $\beta
_{l,r}=0 $ and any $\alpha $. We further discuss the relation between
negative $I_{c}$ and a paramagnetic response.

\section{Basic Equations}

We consider the structure shown in Fig.1. It consists of two superconducting
S and two normal-metal N reservoirs, respectively. They are connected by
ferromagnetic (normal) wires of length, $L$. The superconductors are covered
by thin magnetic layers. These layers are made of weak ferromagnets F$_{l,r}$
whereas the wire between N or S reservoirs consists of a strong ferromagnet F%
$_{st}$ or normal (non-magnetic) metal. The magnetization vector $\mathbf{M}%
_{l,r}=(M\mathbf{n)}_{l,r}$ in the weak ferromagnets are expected to be not
collinear with respect to each other and to the magnetization $\mathbf{M}%
_{st}=M_{st}\mathbf{n}_{z}$ in the magnetic wire so that the LRSTC arises in
the structure due to proximity effect (PE) \cite%
{BVEprl01,BVErev05,EschrigRev11,LinderRev15}. The unit vector $\mathbf{n}$
is characterized by the polar\ ($\beta $) and the azimuthal ($\alpha $)
angles in the usual way $\mathbf{n}=(\sin \beta \cos \alpha $, $\sin \beta
\sin \alpha $, $\cos \beta )$. In particular, non-collinearity means that $%
\beta \neq 0$ and $\alpha_{l} \neq \alpha_{r}$. In principle, for LRSTC
penetration it should not matter whether the F wire is magnetic or
non-magnetic yet in practice possible magnetic inhomogeneities in the F wire
may shorten the penetration length of the LRSTC \cite{IvanovFominSkvorts09}.
In what follows we calculate the conductance between the N reservoirs, $G$,
and its deviation due to the PE from that in the normal state (above $T_{c}$%
) $G_{nor}$. The voltages in the N reservoirs are assumed to be $\pm V$, the
electric potential in the S reservoirs is set to zero and their phases are
different and equal to $\pm \varphi/2 $. In the considered symmetric N -
F(n) - N circuit the electric potential $V$ equals zero in the center of the
cross, \textit{i.e.} $V(y)=0$ at $y=0$, such that there is no voltage
difference between the S-F(n)-S superconducting circuit and the N - F(n) - N
circuit. In this respect, the case considered here differs from that studied
in Ref.\cite{VolkovPRL95}, where a voltage drop between the center of the $x$%
-circuit ($V(x)$ at $x=0$) and the S reservoirs was of the order of $\Delta
/e$. This means that, unlike the current situation, the quasiparticles in
the S-F(n)-S circuit, considered in Ref.\cite{VolkovPRL95} were not in
equilibrium with condensate.

The calculations are carried out on the basis of equations for generalized
quasiclassical Green's functions $\check{G}$ \cite%
{LOrev,VZK93,ZaitsevVolkov99,ZaikinBelzig99,KopninBook}, which are widely
and successfully used in the theory of S/N or S/F structures \cite%
{VZK93,Zaitsev94,NazarovStoof96,Zaikin97,ZaitsevVolkov99}. Here, the
elements of the matrix $\check{G}$ are the retarded (advanced) Green's
functions $\check{g}^{R(A)}=\check{G}_{11,22}$ as well as the Keldysh
function $\check{g}=\check{G}_{12}$, which in turn are also matrices in the
Gor'kov-Nambu ($\hat{\tau}$ matrices) and the spin ($\sigma $ matrices)
space, respectively. In particular, the Keldysh function $\check{g}$ is
written in terms of matrix distribution functions $\check{n}$
\begin{equation}
\check{g}=\check{g}^{R}\cdot \check{n}-\check{n}\cdot \check{g}^{A}
\label{G1}
\end{equation}%
where the matrix $\check{n}$ can be represented as
\begin{equation}
\check{n}=\hat{n}_{od}\cdot \hat{\tau}_{0}+\hat{n}_{ev}\cdot \hat{\tau}_{3}
\label{G1a}
\end{equation}%
where $\hat{n}_{od}$ and $\hat{n}_{ev}$ are matrices in the spin space and $%
\tau _{i}$ are matrices in the Gor'kov-Nambu space. The reservoirs S and N
are supposed to be in equilibrium so that the distribution functions $\hat{n}%
_{od,ev}=\hat{\sigma}n_{odd,ev}$ are equal to
\begin{eqnarray}
n_{od}=\tanh (\epsilon /2T)\text{, }n_{ev}=0; &&\text{S reservoirs,}
\nonumber  \label{nS} \\
n_{od,ev}(L_{y})=F_{\pm }(V); &&\text{N reservoir at the top,}  \label{nN} \\
\hat{n}_{nod,ev}(-L_{y})=F_{\pm }(-V); &&\text{N reservoir at the bottom,}
\nonumber
\end{eqnarray}%
where $F_{\pm }(\epsilon ,V)=\frac{1}{2}[\tanh ((\epsilon +eV)/2T)\pm \tanh
((\epsilon -eV)/2T]$. The subscripts $ev,odd$ denote even (odd) functions of
the energy $\epsilon $, respectively. $\check{g}$, $\check{g}^{R(A)}$ also
obey the normalization condition
\begin{eqnarray}
\check{g}^{R}\cdot \check{g}+\check{g}\cdot \check{g}^{A} &=&0,  \label{G2}
\\
\check{g}^{R(A)}\cdot \check{g}^{R(A)} &=&\check{1}  \label{G2a}
\end{eqnarray}%
In the F wire the matrices $\check{g}$ and $\check{g}^{R(A)}$ satisfy the
generalized Usadel equation \cite{LOrev,VZK93,ZaikinBelzig99,KopninBook}
\begin{widetext}
\begin{eqnarray}
D_{v}\nabla (\check{g}^{R}\cdot \nabla \check{g}+\check{g}\cdot \nabla
\check{g}^{A})+i\epsilon \lbrack \hat{\tau}_{3}\cdot \hat{\sigma}_{0},\check{%
g}]+i\kappa _{F}[\hat{\tau}_{3}\cdot \hat{\sigma}_{3},\check{g}] &=&0%
\text{,}  \label{UsG} \\
D_{v}\nabla (\check{g}\cdot \nabla \check{g})^{R(A)}+i\epsilon \lbrack \hat{%
\tau}_{3}\cdot \hat{\sigma}_{0},\check{g}^{R(A)}]+i\kappa _{F}[\hat{\tau%
}_{3}\cdot \hat{\sigma}_{3},\check{g}^{R(A)}] &=&0\text{ }  \label{UsGr}
\end{eqnarray}
\end{widetext}where $\kappa _{F}=\sqrt{E_{F}sign(\omega )/D_{F}}$ is a
parameter which characterizes a decay of the short-range component in the F
film. In the case

Observe that if the wires connecting S and N reservoirs are non-magnetic (n
metals), the last terms vanishes. The charge current\ density $I$ in a wire
with conductivity $\sigma $ is expressed conventionally in terms of matrices
$\check{G}$ and is a sum of the condensate current $I_{S}$ and the
quasiparticle current $I_{qp}$, $I=I_{S}+I_{qp}$, as follows
\begin{equation}
I=\frac{\sigma }{4e}\int d\epsilon \{\check{g}^{R}\cdot \partial _{x}\check{g%
}+\check{g}\cdot \partial _{x}\check{g}^{A}\}_{3,0}  \label{G3}
\end{equation}%
where $\{(..)\}_{3,0}\equiv $ Tr$\{(\hat{\sigma}_{0}\cdot \hat{\tau}%
_{3})\cdot (..)\}/4$. In the symmetric case, considered here, only $I_{S}$
differs from zero in the $x$-wire. It is proportional to $\hat{n}_{eq}=\hat{%
\sigma}_{0}\tanh (\epsilon /2T)$ and, as follows from Eq.(\ref{G3}), is
expressed in terms of the condensate functions $\check{f}_{\omega }$
\begin{eqnarray}
I_{S} &=&\frac{\sigma _{x}}{4e}\int d\epsilon \{(\check{g}^{R}\cdot \partial
_{x}\check{g}^{R}-\check{g}^{A}\cdot \partial _{x}\check{g}^{A})\hat{n}%
_{eq}\}_{3,0}=  \label{G4} \\
&=&i\pi \sigma _{x}\frac{T}{e}\sum_{\omega \geqslant 0}\{\check{f}_{\omega
}\cdot \partial _{x}\check{f}_{\omega }\}_{3,0}\text{.}  \label{G4a}
\end{eqnarray}%
where $\sigma _{x,y}$ are conductivities in the $x$- and $y$- wires, the
condensate Green's function $\check{f}^{R(A)}$ are defined below (see Eq.(%
\ref{G5})); they\textbf{\ }anticommutes with the matrix $\hat{\tau}_{3}\cdot
\hat{\sigma}_{0}$. Eq.(\ref{G4a}) is identical to Eq.(\ref{G4})\ in the
Matsubara representation with $\omega =\pi T(2n+1)$. Here we represent $%
\check{g}^{R(A)}$ functions as follows
\begin{equation}
\check{g}^{R(A)}=\hat{g}^{R(A)}\hat{\tau}_{3}+\check{f}^{R(A)}  \label{G5}
\end{equation}%
The condensate matrix Green's functions $\check{f}^{R(A)}$ have the form $%
\check{f}^{R(A)}=\hat{\tau}_{\perp }\hat{f}^{R(A}$ with $\hat{\tau}_{\perp
}\sim \hat{\tau}_{1,2}$ (see next section). In the vertical wire there is no
supercurrent and the current $I_{y}$ is carried by quasiparticles. Eq. (\ref%
{G3}) can be written as
\begin{equation}
I_{y}=\frac{\sigma _{y}}{4e}\int d\epsilon \partial _{y}\{\hat{n}%
_{ev}\}_{0}[1-\{\check{g}^{R}\cdot \hat{\tau}_{3}\cdot \check{g}%
^{A}\}_{3,0}]=\frac{\sigma _{y}T}{eL_{y}}\int d\zeta \tilde{J}(\zeta )\text{
}  \label{G6}
\end{equation}%
where $\{\hat{n}_{ev}\}_{0}\equiv (1/2)$Tr$(\hat{\sigma}_{0}\hat{n}_{ev})$
and $\zeta =\epsilon /(2T)$. Using Eq.(\ref{G5}), we can represent the
partial current $J(\epsilon )$ in the following form
\begin{equation}
\tilde{J}(\zeta ,y)\text{ }=\frac{L_{y}}{2}(\partial _{y}n_{0})(1-m(\zeta
,y))  \label{G7}
\end{equation}%
where $n_{0}=\{\hat{n}_{ev}\}_{0}$ and
\begin{equation}
m(\zeta ,y)=\frac{1}{4}\{[\check{f}^{R}(\zeta ,y)-\check{f}^{A}(\zeta
,y)]^{2}\}_{0,0}  \label{G8}
\end{equation}%
Here, we use the approximation $\check{g}^{R(A)}\approx \pm (1+(1/2)(\check{f%
}^{R(A)})^{2})$, which follows from Eq.(\ref{G2a}) in the case of a weak PE,
i. e., $|\check{f}^{R(A)}(y)|\ll 1$. Thus, the product $(\hat{g}^{R}\cdot
\hat{g}^{A})_{0}$ is equal to
\begin{equation}
-(\hat{g}^{R}(\zeta ,y)\cdot \hat{g}^{A}(\zeta ,y))_{0}\cong 1+\{\frac{1}{2}(%
\hat{f}^{R}(\zeta ,y)+\hat{f}^{A}(\zeta ,y))^{2}\}_{0}\text{,}  \label{G7a}
\end{equation}%
Thus, the currents $I_{x,y}$ are given by Eq.(\ref{G4a}) and Eqs.(\ref{G5}-%
\ref{G8}), respectively. In order to evaluate further these currents, we
need to determine the condensate Green's functions $\check{f}^{R(A)}$ in the
$x$- and $y$-wires. This we do in the next section.

\begin{figure}[tbp]
\includegraphics[width=0.8\columnwidth]{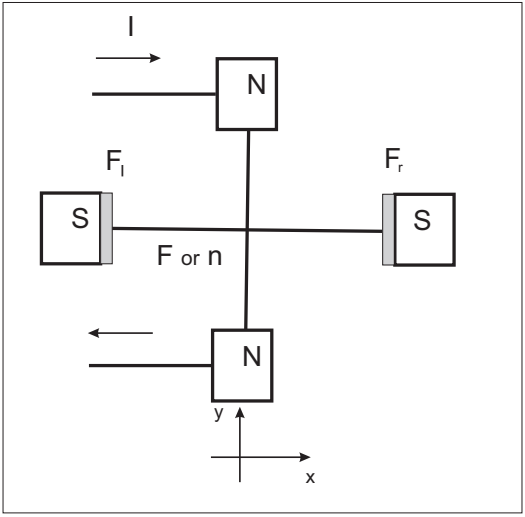}
\caption{(Color online.) Schematic structure of the system under
consideration. The current $I$ is the dissipative current in the vertical
circuit.}
\label{SFnFS}
\end{figure}

\section{Condensate Green's Functions}

\subsection{Condensate Functions in the x-Wire}

In order to simplify the calculation, we assume that the interface
resistance of the cross is larger than the resistance of the F$_{x,y}$ or n$%
_{x,y}$ wires, i. e.
\begin{equation}
R_{B}/w_{x,y}\gg L_{x,y}/\sigma _{x,y}\text{,}  \label{Co0}
\end{equation}%
where $R_{B}$ is the resistance of the interface between $x$ - and $y$-wires
per unit area, $w_{x,y}$ is the width of these wires. This assumption means
that when determining the condensate function in the $x$-wire, we can
neglect the leakage of Cooper pairs from the $x$-wire into the $y$-wire. On
the other hand, the condensate into the $y$-wire is determined by a small
leakage from the $x$ wire. The generalization for the case of arbitrary $%
R_{B}$ is straightforward and does not change the results qualitatively.

In what follows we evaluate the condensate function $\check{f}_{\omega
}(x,y) $ in the $x$- and $y$-wires. In the SF$_{l}$ - n(F) - F$_{r}$S wire,
the condensate functions in the Matsubara representation $\check{f}_{\omega
}(x,0)$ obey the linearized Usadel equation, Eq.(\ref{UsGr})

\begin{widetext}
\begin{equation}
-\partial _{xx}^{2}\check{f}_{\omega }+\kappa _{\omega }^{2}\check{f}%
_{\omega }+i(\kappa _{F}^{2}/2)[\hat{\sigma}_{3},\check{f}_{\omega
}]_{+}+i(\kappa _{l}/2)\delta (x+L_{x})[\hat{\sigma}_{1},\check{f}_{\omega
}]_{+}+i(\kappa _{r}/2)\delta (x-L_{x})[\hat{\sigma}_{r},\check{f}_{\omega
}]_{+}=0\text{,}  \label{Co1}
\end{equation}
\end{widetext}where $\kappa _{\omega }^{2}=2|\omega |/D_{F}$, $\kappa
_{F}^{2}=E_{F}$sign $(\omega )/D_{F}$, $\kappa _{l,r}=(wE)_{l,r}\mbox{sign}%
(\omega )/D_{F}$. Here, $E_{H}$ and $E_{l,r}$ are the exchange energy in the
strong or left (right) ferromagnetic films, respectively, $w$ is the
thickness of the F$_{l,r}$ films. The matrices $\hat{\sigma}_{l,r}\equiv (%
\mathbf{\hat{\sigma}n})_{l,r}$are defined in the following way
\begin{equation}
\hat{\sigma}_{l,r}=\{(\hat{\sigma}_{1}\cos \alpha +\hat{\sigma}_{2}\sin
\alpha )\sin \beta +\hat{\sigma}_{3}\cos \beta \}_{l,r}.  \label{Co1a}
\end{equation}%
The unit vector $\mathbf{n}$ has the components $\mathbf{n}_{l,r}=(\cos
\alpha \sin \beta $,$\sin \alpha \sin \beta $,$\cos \beta )_{l,r}$. It is
important to note that the spin and the orbital degree of freedom are
decoupled in our model as no spin-orbit interaction is included. Therefore,
the components of the vectors $\mathbf{n}_{l,r}=(n_{x},n_{y}$,$n_{z})$ are
arbitrarily oriented independent of the coordinate system shown in Fig.1. In
particular, we would like to stress that the magnetisation vector\textbf{\ }$%
\mathbf{M}=M_{0}\mathbf{n}$ is not necessarily oriented along the $x$-axis
shown in Fig.1 if $\alpha =0$ and $\beta =\pi /2$. The $\delta $-functions
in Eq.(\ref{Co1}) refer to the thickness of the F$_{l,r}$ layers, $w_{l,r}$,
which is assumed to be is thinner than $\kappa _{l,r}^{-1}$. In addition,
Eq.(\ref{Co1}) is supplemented by boundary conditions \cite%
{ZaitsevBC,KupLukichev88}
\begin{equation}
\partial _{x}|\check{f}_{\pm L}=\pm \kappa _{b}F_{S}\hat{\tau}_{r,l}\cdot
\hat{\sigma}_{0}\text{.}  \label{Co2}
\end{equation}%
where $\kappa _{b}=1/(R_{b}\sigma _{x})$, $R_{b}$ and $\sigma _{x}$ are the
S/F interface resistance (per unit area) and the conductivity of the $x$%
-wire. The matrices $\hat{\tau}_{l,r}$ are defined as follows

\begin{equation}
\hat{\tau}_{l,r}=\hat{\tau}_{1}(\cos (\varphi /2)\pm i\hat{\tau}_{3}\sin
(\varphi /2))  \label{Co2b}
\end{equation}%
The Green's functions $F_{S}$ have a standard BCS form: $F_{S}=\Delta /\sqrt{%
\omega ^{2}+\Delta ^{2}}$. The quantities $\pm \varphi /2$ are the phases of
the order parameter in the right (left) superconductors.

By integrating Eq.(\ref{Co1}) over $x$ in the vicinity of the left (right) SF%
$_{l,r}$ interfaces, we can get rid of the $\delta $-functions from this
equation and obtain new effective BCs for $\partial _{x}\check{f}$
\begin{eqnarray}
\lefteqn{\partial _{x}\check{f}_{\omega }(\pm L,0)=}  \nonumber \\
&&\pm \{\kappa _{b}F_{S}\hat{\tau}_{r,l}\cdot \hat{\sigma}_{0}+i(\kappa
_{r,l}/2)[\hat{\sigma}_{r},\check{f}(\pm L,0)]_{+}\}\text{.}  \label{Co3a}
\end{eqnarray}%
Finally, in order to find the function $\check{f}_{\omega }(x)$ in the SF$%
_{l}$ - F - F$_{r}$S circuit, one has to solve the following equation
\begin{equation}
-\partial _{xx}^{2}\check{f}_{\omega }+\kappa _{\omega }^{2}\check{f}%
_{\omega }+i\kappa _{F}^{2}[\hat{\sigma}_{3},\check{f}_{\omega }]_{+}=0\text{%
,}  \label{Co2a}
\end{equation}%
with the boundary condition (\ref{Co3a}). In the case of the SF$_{l}$ - n - F%
$_{r}$S circuit, the third term should be dropped.

For simplicity we assume that the distance between F$_{l}$ and F$_{r}$, 2L$%
_{x}$, is larger than $\xi _{T}=\sqrt{D_{F}/2\pi T}$. Then, a solution $%
\check{f}_{\omega }(x,0)$ can be written as a sum
\begin{equation}
\check{f}_{\omega }(x,0)=\check{f}_{l}(x,0)+\check{f}_{r}(x,0)  \label{Co4}
\end{equation}%
where the functions $\check{f}_{l,r}(x,0)\equiv $ $\check{f}_{l,r}(x,y|_{0})$
decay exponentially from the left (right) superconductors. We discuss them
for the cases of SF$_{l}$ - F - F$_{r}$S and SF$_{l}$ - n - F$_{r}$S
structures below.

\textit{1) SF$_{l}$ - F - F$_{r}$S structure:} A solution for the case of
the SF$_{l}$ - F - F$_{r}$S circuit has the form

\begin{widetext}
\begin{equation}
\check{f}_{l,r}(x,0)=\hat{\tau}_{l,r}\cdot \sum_{s=\pm }\left[(\hat{\sigma}
_{0}A_{s}+\hat{\sigma}_{3}B_{s})_{l,r}\exp (-\kappa _{s}(L\pm x)+C_{l,r}((%
\mathbf{\hat{\sigma}n})_{l,r}-\hat{\sigma}_{3}n_{z})\exp (-\kappa _{\omega
}(L\pm x))\right]  \label{Co5}
\end{equation}
\end{widetext}where $\kappa _{\pm }^{2}=\kappa _{\omega }^{2}\pm i\kappa
_{F}^{2}$.and the matrices $\hat{\tau}_{l,r}$ are defined in Eq.(\ref{Co2}).
Note, the presence of the term $\hat{\sigma}_{3}n_{z}$ means that only
triplet component with non-collinear spin directions penetrates the F wire
over the length $\kappa _{\omega }^{-1}$.

The constants $A$ and $B$ which characterize the singlet and $B_{l\pm }$
triplet short-range components are equal to $A_{l-}=A_{l+}$, $B_{l+}=A_{l+}$%
, $B_{l-}=-A_{l-}$, $A_{l+}=(\kappa _{b}F_{S}-i\kappa _{l}C_{l})/2\kappa
_{+}=(\kappa _{-}/\kappa _{l})A$. The amplitude of the LRSTC $C$, which we
are mostly interested is
\begin{equation}
C_{l}=-i\frac{\kappa _{b}\kappa _{l}\text{Re}\kappa _{+}}{\kappa _{l}^{2}%
\text{Re}\kappa _{+}+\kappa _{\omega }|\kappa _{+}|^{2}}F_{S}.  \label{Co6b}
\end{equation}%
The coefficients $\check{f}_{r}(x,0)$ are equal to those in Eqs. (\ref{Co5}-%
\ref{Co6b}) upon replacing $l\Rightarrow r$. The constants $A_{l\pm }$
(singlet) and $B_{l\pm }$(triplet) are the amplitudes of the short-range
components of the condensate. They decay over the length $\xi _{F}\cong
\kappa _{F}^{-1}$, which is much shorter than the length $\xi _{T}=\kappa
_{\omega }^{-1}\cong \sqrt{D_{F}/2\pi T}$ in the case of a strong
ferromagnet F ($T,\Delta \ll E_{F}$). The last term in Eq.(\ref{Co5}) refers
to the LRSTC. It penetrates the F wire on the distance of the order of $\xi
_{T}$.

\textit{2) SF$_{l}$- n - F$_{r}$S structure:} Here, the solution is given by
\begin{equation}
\check{f}_{l,r}(x,0)=\hat{\tau}_{l,r}\cdot \{a_{l,r}\hat{\sigma}_{0}+C_{l,r}%
\hat{\sigma}_{l})\exp (-\kappa _{\omega }(L_{x}\pm x))  \label{Co7}
\end{equation}%
The coefficients $a_{l}$ and $a_{l}$ can be found from the boundary
condition (\ref{Co3a})
\begin{equation}
a_{l,r}=\frac{\kappa _{b}\kappa _{\omega }}{\kappa _{l,r}^{2}+\kappa
_{\omega }^{2}}F_{S}\text{, }C_{l,r}=-i\frac{\kappa _{b}\kappa _{l,r}}{%
\kappa _{l,r}^{2}+\kappa _{\omega }^{2}}F_{S}  \label{Co8}
\end{equation}%
In this case, both components, singlet and triplet, decrease over a long
distance of the order $\xi _{T}$.

\subsection{Condensate Functions in the y-Wire}

To find the condensate function in the $y$-wire induced by PE we assume that
the widths of the wire $w_{x,y}$ are less than $\xi _{T}$. Then one can
write Eq.(\ref{Co2a}) for the LRSTC in $y$-wire as follows
\begin{equation}
-\partial _{yy}^{2}\check{f}_{\omega }(0,y)+\kappa _{\omega }^{2}\check{f}%
_{\omega }(0,y)=\kappa _{B}^{2}w_{x}\check{f}_{\omega }(0,0)\delta (y)\text{%
, }  \label{Co9}
\end{equation}%
where the term on the $r.h.s$ is a source of the Cooper pairs leaking from
the $x$-wire. The coefficient $\kappa _{B}=1/(R_{B}\sigma _{y})$ is related
to the interface resistance F$_{x}$/F$_{y}$ (or n$_{x}$/n$_{y}$) per unit
area. The contact of the $y$-wire with the N reservoirs is supposed to be
ideal so that the boundary conditions for the $\check{f}_{\omega }(0,y)$
function is $\check{f}_{\omega }(0,\pm L_{y})=0$. Then the solution to Eq.(%
\ref{Co9}) satisfying this boundary condition is given by
\begin{equation}
\check{f}_{\omega }(0,y)=\frac{\kappa _{B}^{2}w_{x}}{2\kappa _{\omega }}%
\frac{\sinh (\kappa _{\omega }(L_{y}-|y|))}{\cosh \theta _{\omega y}}\check{f%
}_{\omega }(0,0)\text{, }  \label{Co11}
\end{equation}%
where $\theta _{\omega y}=\kappa _{\omega }L_{y}$, and $\check{f}_{\omega
}(0,0)$ is given by Eqs.(\ref{Co5}-\ref{Co7}) and can be expressed as
\begin{equation}
\check{f}_{\omega }(0,0)=\Bigg\{%
\begin{array}{l}
\sum_{s=l,r}\hat{\tau}_{s}\cdot (\hat{\sigma}_{s}-\hat{\sigma}%
_{3}n_{z})C_{s}\exp (-\kappa _{\omega }L),\text{\textit{case 1}} \\
\sum_{s=l,r}\hat{\tau}_{s}\cdot (a_{s}\hat{\sigma}_{0}+C_{s}\hat{\sigma}%
_{s})\exp (-\kappa _{\omega }L),\text{\textit{case 2}}%
\end{array}
\label{Co12}
\end{equation}%
Knowing the condensate functions, we find the Josephson current $I_{J}$
between the superconductors S and corrections to the conductance between the
N reservoirs due to the PE.

\section{Conductance of the y-Wire}

In this section we calculate the conductance of the $y$-wire. If the
condition, Eq.(\ref{Co0}), is fulfilled one can neglect the leakage of the
current $I_{y}$ into the $x$-wire and use the Eqs.(\ref{G6})-(\ref{G8}). The
partial current $J(\epsilon )$ in Eq.(\ref{G6}) does not depend on the $y$%
-coordinate as can be seen from taking the trace of Eq.(\ref{UsG})
multiplied by the matrix $\hat{\sigma}_{0}\cdot \hat{\tau}_{3}$. Thus, we
have
\begin{equation}
D_{y}\partial _{y}\{\check{g}^{R}\cdot \partial _{y}\check{g}+\check{g}\cdot
\partial _{y}\check{g}^{A}\}_{3,0}\equiv D_{y}\partial _{y}\tilde{J}%
(\epsilon )=0\text{, }  \label{C1}
\end{equation}%
Since the N/F(n) contacts are assumed to be ideal, the distribution function
$n_{0}(\pm L_{y})$ should coincide with the distribution functions $F_{\pm
V} $ in the N reservoirs, i. e. $n_{0}(\pm L_{y})=F_{\pm V}$, where $F_{\pm
V}(\epsilon )$ are defined in Eq.(\ref{nN}). From Eq.(\ref{G7}) we find the
partial current (see \cite{VZK93})
\begin{eqnarray}
\tilde{J}(\zeta )\text{ } &=&\frac{F_{V}(\zeta )}{1+\langle m(\epsilon
,y)\rangle }\cong  \nonumber \\
&\cong &F_{V}(\zeta )(1-\langle m(\zeta ,y)\rangle )  \label{C3a}
\end{eqnarray}%
where we used the smallness of the condensate functions.\ The distribution
function in the upper N reservoir $F_{V}(\zeta )$ is $F_{V}(\zeta
)=(1/2)[\tanh (\zeta +v)-\tanh (\zeta -v)]$ with $v=eV/2T$. The function $%
\langle m(\epsilon ,y)\rangle =(1/L_{y})\int_{0}^{L_{y}}dy\{m(\epsilon ,y)\}$
can be expressed as
\begin{widetext}
\begin{equation}
\langle m(\zeta ,y)\rangle =\frac{1}{4}\langle \{(\check{f}^{R}(\zeta,0,y))^{2}+(%
\check{f}^{A}(\zeta,0,y))^{2}-2\check{f}^{R}(\zeta,0,y)\cdot \check{f}%
^{A}(\zeta,0,y)\}_{0,0}\rangle  \label{C3a}
\end{equation}%
\end{widetext}According to Eq.(\ref{G6}), the normalized correction to the
current $\delta \tilde{I}_{y}\equiv \delta IeL_{y}/(2T\sigma _{y})$ caused
by PE is
\begin{equation}
\delta \tilde{I}_{y}\equiv \delta IeL_{y}/(2T\sigma _{y})=\frac{1}{2}%
\int_{-\infty }^{\infty }d\zeta F_{V}(\zeta )\langle m(\zeta ,y)\rangle
\label{C4}
\end{equation}%
where $\check{f}^{R(A)}$ is defined in Eq.(\ref{Co11}). The average $\langle
\delta m(\zeta ,y)\rangle $ is easily found with the help of Eqs.(\ref{Co11}%
)-(\ref{Co12}). In particular, we represent $\langle m(\epsilon ,y)\rangle $
in the form
\begin{equation}
\langle m(\zeta ,y)\rangle =m^{RR}(\zeta )+m^{AA}(\zeta )-2m^{RA}(\zeta )
\label{C5}
\end{equation}%
where
\begin{eqnarray}
m^{RR}(\zeta ) &=&\frac{1}{4}\langle \{(\check{f}^{R}(\zeta
,0,y))^{2}\}_{0,0}\rangle  \label{C5a} \\
m^{RA}(\zeta ) &=&\frac{1}{4}\langle \{\check{f}^{R}(\zeta ,0,y)\cdot \check{%
f}^{A}(\zeta ,0,y)\}_{0,0}\rangle  \label{C5b}
\end{eqnarray}

\begin{figure}[tbp]
\includegraphics[width=0.8\columnwidth]{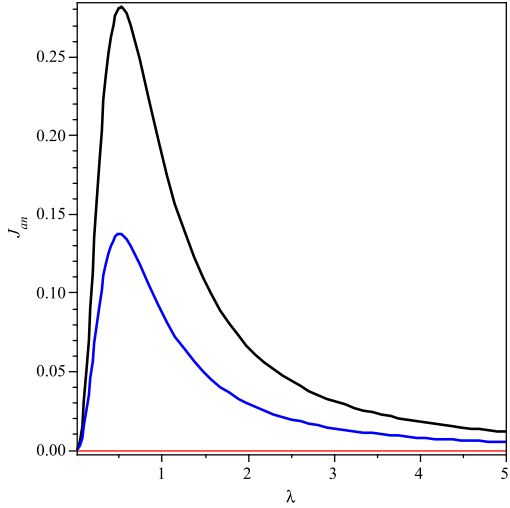}
\caption{(Color online.) The normalized amplitude of the oscillating part of
the current $J_{an}=$ $\mathcal{I}_{an}(P_{x},P_{y},v,\protect\lambda )/%
\mathcal{I}_{c,n}(P_{x})$ as a function of the parameter $\protect\lambda $
for $P_{x}=1$(black), and $P_{x}=2$(blue) with different scaling factors: $%
1\ast J_{an}(1) $ and $0.02\ast J_{an}(2)$ . Other parameters are: $P_{y}=5$%
, $v=1$. }
\label{Fig2}
\end{figure}

\begin{figure}[tbp]
\includegraphics[width=0.8\columnwidth]{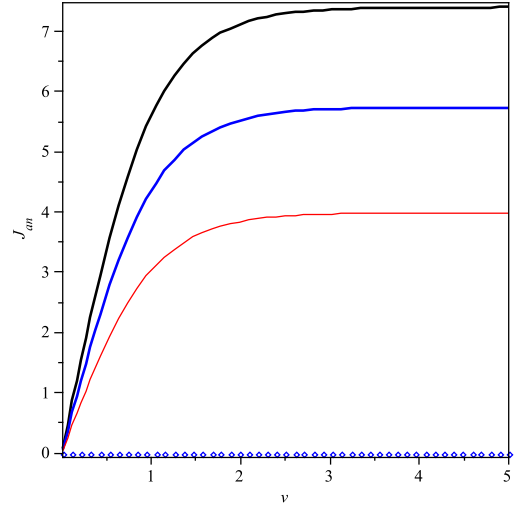}
\caption{(Color online.) The same quantity as in Fig.2 as a function of the
dimensionless voltage $v=eV/2T$ for $P_{x}=1$ (black), $P_{x}=2$ (blue) and $%
P_{x}=3$ (red). The scaling factors are: $30\ast J_{an}(1)$, $1\ast
J_{an}(2) $ and $0.03\ast J_{an}(3)$.}
\label{Fig3}
\end{figure}

\begin{figure}[tbp]
\includegraphics[width=0.8\columnwidth]{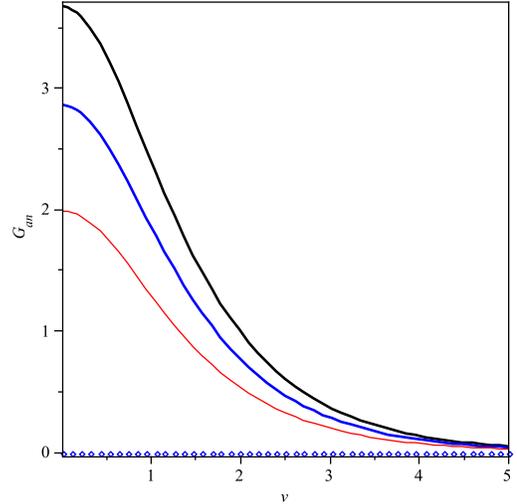}
\caption{(Color online.) The normalized differential conductance $G_{an}$ vs
the dimensionless voltage $v=eV/2T$ for the same parameters as in Fig.3.}
\label{Fig4}
\end{figure}

The terms $m^{RR}(\zeta )$ and $m^{AA}(\zeta )$ contribute to the so-called
regular part of the current $\delta \tilde{I}_{y}$
\begin{equation}
\delta \tilde{I}_{reg}=-\frac{1}{2}\int_{-\infty }^{\infty }d\zeta
F_{V}(\zeta )[m^{RR}(\zeta )+m^{AA}(\zeta )]\text{.}  \label{Reg1}
\end{equation}%
The anomalous current is given by

\begin{equation}
\delta \tilde{I}_{an}=2\int_{0}^{\infty }d\zeta F_{V}(\zeta )m^{RA}(\zeta )%
\text{.}  \label{Anom}
\end{equation}%
The integral in Eq.(\ref{Reg1}) can be transformed into the sum over
Matsubara frequencies
\begin{equation}
\delta \tilde{I}_{reg}=-2\pi \text{Im}\sum_{n\geqslant 0}m(\zeta _{n}+2iv)
\label{Reg2}
\end{equation}%
where $m(\zeta _{n})=m^{RR}(\epsilon _{n}/2T)$, $\epsilon _{n}=i\omega
_{n}=T\zeta _{n}$, $\zeta _{n}=\pi (2n+1)$.

\begin{figure}[tbp]
\includegraphics[width=0.8\columnwidth]{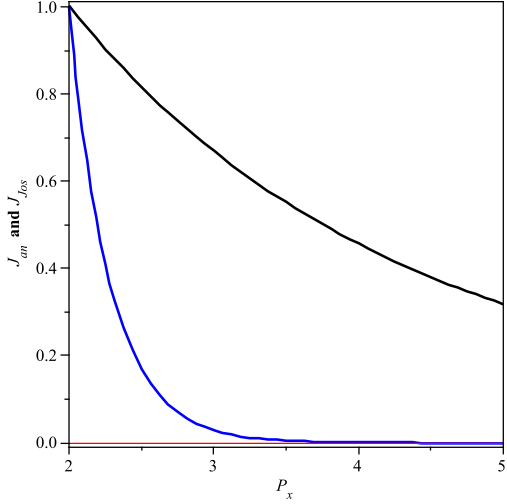}
\caption{(Color online.) Comparison of the amplitude of oscillatory part of
the current $J_{an}=0.23\ast \mathcal{I}_{an}(P_{x})/\mathcal{I}_{c,n}(2)$
and the Josephson critical current $J_{Jos}=\mathcal{I}_{c,n}(P_{x})/%
\mathcal{I}_{c,n}(2)$ in S - n - S junction as functions of the parameter $%
P_{x}$. One can see that the phase-coherent part of the current in N - F(n)
- N circuit is much larger than the Josephson critical current $I_{c}$ in S
- n - S junction.}
\label{Fig5}
\end{figure}

We need to evaluate $m^{RA}(\epsilon )\equiv \langle \{\check{f}^{R}(y)\cdot
\check{f}^{A}(y)\}_{0,0}\rangle $ as $m^{RR}(\epsilon )$ and $%
m^{AA}(\epsilon )$ can be found directly from it. The function $%
m^{RA}(\epsilon )$ can be obtained with the aid of Eq.(\ref{Co12}). In
particular, we find

\begin{widetext}
\begin{equation}
m^{RA}(\zeta )=s^{RA}(\zeta)\exp (-(\theta _{x}^{R}+\theta
_{x}^{A})\sum_{s=l,r}[C_{s}^{R}C_{s}^{A}+C_{s}^{R}C_{\bar{s}}^{A}\chi_{1}
(\alpha,\beta)\cos \varphi ], \textit{\ case 1} \label{C6a}
\end{equation}
and
\begin{equation}
m^{RA}(\zeta )=s^{RA}(\zeta
)\exp(-(\theta _{x}^{R}+\theta _{x}^{A})\sum_{s=l,r}[a_{s}^{R}a_{s}^{A}+C_{s}^{R}C_{s}^{A}+\cos \varphi
(a_{s}^{R}a_{\bar{s}}^{A}+\chi_{2} (\alpha,\beta)C_{s}^{R}C_{\bar{s}}^{A})], \textit{\ case 2} \label{C6b}
\end{equation}%
\end{widetext}with subscripts $s=l,r$ and $\bar{s}=r,l$. The coefficients $a$
and $C$ are also functions of $\zeta $. The function $m^{RR}(\epsilon )$ is
obtained from Eqs.(\ref{C6a}-\ref{C6b}) by replacing $A\Rightarrow R$ and
changing its sign. The angle-dependent function $\chi _{1,2}(\alpha ,\beta )$
is then determined as

\begin{equation}
\chi (\alpha ,\beta )=\Bigg\{%
\begin{array}{l}
\chi _{1}(\alpha ,\beta )\text{,}\quad \quad \quad \mathit{\quad \ \ \ \ \ \
\ \ \ \ }\text{\textit{case 1}} \\
\chi _{1}(\alpha ,\beta )+\cos \beta _{l}\cos \beta _{r}\text{,\ \ \ \textit{%
case 2}}%
\end{array}
\label{J3Ang}
\end{equation}%
where $\chi _{1}(\alpha ,\beta )=\cos (\alpha _{r}-\alpha _{l})\sin \beta
_{l}\sin \beta _{r}$. The angles $\alpha _{l,r}$ and $\beta _{l,r}$
determine the orientation of the unit vector $\mathbf{n}$, see Eq.(18). The
coefficients $s^{RA}(\zeta )$ and $s^{RR}(\zeta )$ are then equal to
\begin{widetext}
\begin{eqnarray}
s^{RA}(\epsilon ) &=&\frac{\theta _{B}^{2}}{16|\theta _{y}^{R}(\zeta
)|^{2}}\frac{\text{Im}[\theta _{y}(\zeta )\tanh \theta _{y}^{\ast }(\zeta )]%
}{\text{Re}\theta _{y}(\zeta )\text{Im}\theta _{y}^{\ast }(\zeta )}\text{,}
\label{C7} \\
\text{ }s^{RR}(\epsilon _{n}) &=&\frac{\theta _{B}^{2}}{16\theta
_{y}^{2}(\epsilon )\cosh ^{2}\theta _{y}(\epsilon )}[\frac{\sinh (2\theta
_{y}(\epsilon ))}{2\theta _{y}(\epsilon )}-1]\text{,}
\label{C7a}
\end{eqnarray}%
\end{widetext}where $\theta _{B}=(\kappa _{B}^{2}w)L_{y}$, $\theta
_{x,y}(\zeta )=P_{x,y}\sqrt{\zeta }$, $P_{x,y}=\sqrt{2T/E_{x,y}}$, $%
E_{x,y}=\{D/L^{2}\}_{x,y}$.

The most interesting parts of the current $\delta \tilde{I}=\delta \tilde{I}%
_{reg}+\delta \tilde{I}_{an}$ are the parts which depend on the phase $%
\varphi $ and angles $\{\alpha $, $\beta \}$. We represent them in the form
\begin{eqnarray}
\delta \tilde{I}_{reg} &=&p_{i}\mathcal{I}_{reg}(v)\cos \varphi \chi
_{i}(\alpha ,\beta )\text{.}  \label{C8} \\
\delta \tilde{I}_{an} &=&p_{i}\mathcal{I}_{an}(v)\cos \varphi \chi
_{i}(\alpha ,\beta )\text{.}  \label{C8a}
\end{eqnarray}%
where the subindex $i=1,2$ stands for the \textit{cases 1,2}. The amplitudes
$\mathcal{I}_{reg}$, $\mathcal{I}_{an}$ are given by Eq.(\ref{A1}-\ref{Ap})
in Appendix.

Eqs.(\ref{C8}-\ref{C8a}) describe the oscillating part $\delta \tilde{I}$ of
the current in the $y$-wire. It turns out that the function $\mathcal{I}%
_{reg}$,is much less than $\mathcal{I}_{an}$: $\mathcal{I}_{reg}$/$\mathcal{I%
}_{an}\lesssim 10^{-3}$ for $P_{x}=2$, $P_{y}=1$, and $\lambda _{1}=0.5$.
One can show that, with increasing $P_{x}$, the anomalous part decays slower
than the regular part (see Fig.5). Whereas the regular part decays with $%
P_{x}$ exponentially, $\mathcal{I}_{reg}\sim \exp [-2P_{x}(\pi
^{2}+4v^{2})^{1/4}]$, the anomalous part $\mathcal{I}_{an}$ decreases in a
power-law fashion. Earlier the slow decrease of anomalous contribution in
space has been obtained in other problems \cite%
{Lempitsky98,TakayanVolkovPRL96}.

In Figs.2-5 we plot the dependence the normalized current $J_{an}=\mathcal{I}%
_{an}/\mathcal{I}_{J,n}$ and the differential conductance $G_{an}=(d\mathcal{%
I}_{an}(v)/dv)/\mathcal{I}_{J,n}$ vs different variables, i. e., vs the
''voltage $v$, the parameter $\lambda $. We plot Fig.2 for $\lambda =\lambda
_{1}$; qualitatively similar form has the curve for$\lambda =\lambda _{2}$.
The current $J_{an}(\lambda )$ increases from zero (no LRSTC in the absence
of ferromagnetic films F$_{l,r}$ with non-collinear magnetisations, i. e.,
at $\lambda =0$), reaches a maximum and then decreases to zero at large $%
\lambda $. As a function of the normalized voltage $v$ the current $%
J_{an}(v) $ increases to a constant value whereas the differential
conductance $G_{an}$ drops to zero. The corresponding curves are shown in
Figs.3 and 4 for different values of the parameter $P_{x}$; $P_{x}=1$, $2$
and $3$ from top to bottom.

\section{Josephson Current}

In this section we calculate the Josephson current in an SF$_{l}$/F$_{st}$/F$%
_{r}$S \ and SF$_{l}$/n/F$_{r}$S junctions using formulas for the condensate
functions (see Eqs.(\ref{Co5}-\ref{Co8})). Note that the obtained formulas
for the Josephon current are also applicable to fully planar structures. The
Josephson current in magnetic junctions was calculated in many theoretical
papers. Ballistic regime was considered in Refs. \cite%
{HaltermanPRB15,RadovicPRB10,Mel'nikovPRL12} and diffusive case was analyzed
in many papers for equilibrium \cite%
{BVEprl03,FominovPRB05,AnischPRB06,EschrigPRB07,BraudePRL07,HouzetPRB07,EschrigPRL08,EschrigPRL09,Kawabata10,GolubovPRL10,HaltermanSST16}
and nonequilibrium cases \cite%
{Bobkov10,LinderHaltermanPRB14,Silaev19,Rahmonov19}. Since we assume that
the length between superconductors $2L_{x}$ is larger than $\xi _{F}=\sqrt{%
D_{F}/E_{F}}$, we need to take into account only the LRSTC, i. e., the
latter term in Eq.(\ref{Co5}) and both components in Eq.(\ref{Co7}).
Substituting these components in Eq.(\ref{G4a}), we obtain
\begin{equation}
I_{J}=I_{c}(\alpha ,\beta )\sin \varphi \text{,}  \label{J1}
\end{equation}%
where $\varphi $ is the phase difference and the critical current $%
I_{c}=I_{c}(\alpha ,\beta )$ depends on orientation of the magnetization
vectors $\mathbf{M}_{l,r}$ in the left and right layers F$_{l,r}$. This
dependence has different forms in the cases \textit{1} and \textit{2}.

The critical current $I_{c}$ is equal to

\begin{widetext}
\begin{eqnarray}
I_{c}(\alpha ,\beta ) &=&-(4\pi T/e)\sigma _{x}\chi_{1} (\alpha ,\beta
)\sum_{\omega }|C_{r}C_{l}|\kappa _{\omega }\exp (-2\kappa _{\omega }L_{x}),%
\text{\textit {Case 1}}  \label{J2} \\
I_{c}(\alpha ,\beta ) &=&(4\pi T/e)\sigma _{x}\sum_{\omega }[a_{l}a_{r}-\chi_{2}
(\alpha ,\beta )C_{r}C_{l}]\kappa _{\omega }\exp (-2\kappa _{\omega }L), \text{%
\textit{Case 2}}  \label{J2a}
\end{eqnarray}%
\end{widetext}where the coefficients $C_{l,r}$ are defined in Eqs.(\ref{Co6b}%
) and the function $\chi (\alpha ,\beta )$ in Eq.(\ref{J3Ang}). The
coefficients $a_{l,r}$ and $C_{l,r}$ are given in Eq.(\ref{Co8}).

Note, at $\beta _{l}=\beta _{r}=\pi /2$, the sign of the critical current $%
I_{c}$ is determined by the difference $(\alpha _{r}-\alpha _{l})$. If this
difference is equal to $\pi $, that is, the vector $\mathbf{n}_{\perp }$
rotates by $\pi $ over the length $2L$, then $I_{c}$ is positive. If the
rotation angle is zero, the current $I_{c}$ becomes negative. The first case
can be called topological since the winding number of the vector $\mathbf{n}%
_{\perp }$ in the first case is $var\{angle(\mathbf{n}_{\perp })\}/\pi =1$,
while in the second case $var\{angle(\mathbf{n}_{\perp })\}/\pi =0$. The
sign change of the current $I_{c}$ can occur not only in magnetic Josephson
junctions, but also in multiterminal structures with a non-equilibrium
distribution function \cite{VolkovPRL95,ZaikinPRL98,Yip98, Bobkov10,Bobkov12}%
. In other words, the SF$_{l}$-F-F$_{r}$S circuit models a ferromagnetic
wire with two domain walls. The topological configuration with $(\alpha
_{l}-\alpha _{r})=\pi $ corresponds to a positive $I_{c}$ (the magnetization
vector $\mathbf{M}$ rotates clockwise or counter clockwise) that the
critical current $I_{c}(\alpha )$ is positive if the difference $(\alpha
_{l}-\alpha _{r})=\pi $. In a non-topological case when the vector $\mathbf{M%
}$ rotates first by $\pi /2$ from $\alpha _{l}=0$ at F$_{l}$ and then
returns to $\alpha _{r}=0$, the critical current in the last case is
negative.

The first term in Eq.(\ref{J2a}) is due to the singlet component. The second
term that changes sign by varying the angles $\alpha $ is caused by the
triplet component. If the parameter $\kappa _{l,r}$ is small compared to $%
\kappa _{\omega }$, i. e. $\kappa _{l,r}\xi _{T}\ll 1$, then the first term
in square brackets dominates and the critical current is positive. In the
opposite limit, $\kappa _{l,r}\xi _{T}\gg 1$, the second term in Eq.(\ref%
{J2a}) is larger than the first one and the sign of $I_{c}$ depends on
orientations of the vector $\mathbf{M}_{l,r}$.

\bigskip In analogy with Eqs.(\ref{C8}-\ref{C8a}), the angle-dependent part
of the critical current $I_{c}(\alpha ,\beta )$ can be written as

\begin{eqnarray}
\tilde{I}_{c}(\alpha ,\beta ) &=&p_{J}\mathcal{I}_{Jos}\chi (\alpha ,\beta )%
\text{,}  \label{J3} \\
\mathcal{I}_{Jos} &=&2\pi \sum_{n\geqslant 0}|C(\zeta _{n})|^{2}\sqrt{\zeta
_{n}}\exp (-2P_{x}\sqrt{\zeta _{n}})  \label{J3a}
\end{eqnarray}%
where $p_{J1,2}$ are given in Appendix (Eqs.(\ref{ApJ})).

In order to compare the formulas for the currents $I_{an}$, $I_{Jos}$, it is
useful to write down the formula for the critical current $I_{c,n}$ in an S
- n - S junction. The formula for $I_{c,n}$ can be directly found from Eq.(%
\ref{J2a}) by setting $C=0$

\begin{equation}
I_{c,n}=2\pi (\frac{\kappa _{b}}{\kappa _{T}})^{2}\sum \frac{\exp (-2P_{x}%
\sqrt{\zeta _{n}})}{\sqrt{\zeta _{n}}}\frac{\tilde{\Delta}^{2}}{\tilde{\Delta%
}^{2}+\zeta _{n}^{2}}\text{.}  \label{J4}
\end{equation}%
where $\tilde{\Delta}=\Delta /(2T)$.

At $\varphi =0$, the Josephson current $I_{J}$ turns to zero for any angles $%
\alpha $ and $\beta $. In the terminology of Ref.\cite{MoorVE15}, the
obtained result corresponds to the nematic case in contrast to the
ferromagnetic one when the Josephson current $I_{J}\neq 0$ even for $\varphi
=0$. The phase-current relation in the latter case has the form $%
I_{J}=I_{c}\sin (\varphi +\psi )$, where $\psi $ is an angle dependent
constant. The unusual phase dependence of the critical Josephson current may
arise in the presence of spin-orbit interaction \cite%
{Krive05,Buzdin08,Balseiro08,Nazarov14,TokatlyPRB15,Silaev17}, in the case
of spin filters \cite{BuzdinAPL20,MoorVE15} or in S/AF/S Josephson junctions
with antiferromagnetic (AF) layer \cite{RabinovichPRR19}.

\begin{figure}[tbp]
\includegraphics[width=0.8\columnwidth]{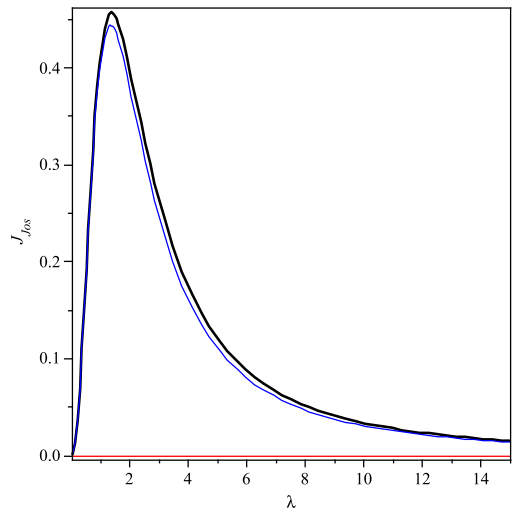}
\caption{(Color online.) The ratio of the critical Josephson current in the
system under consideration and in a S - n - S Josephson junction, $J_{Jos}=$
$\mathcal{I}_{Jos}(P_{x},\protect\lambda )/\mathcal{I}_{c,n}(P_{x})$, with
equal S/n or S/F$_{l,r}$ interface transparency as a function of the
parameter $\protect\lambda $ for $P_{x}=1$ (black) and $P_{x}=2$ (blue).}
\label{Fig6}
\end{figure}

In the considered nematic case, the angle dependence of the current $I_{c}$
is determined by the function $\chi (\alpha ,\beta )$. For $\beta _{l}=\beta
_{r}=\pi /2$ and $\alpha _{l}=-\alpha _{r}=\alpha $, the angle dependence of
$I_{c}$, Eq.(\ref{J4}), coincides with that obtained by Braude and Nazarov
(see Eq.(8) for the critical current $I_{c}=I_{\uparrow }+I_{\downarrow }$
in Ref.\cite{BraudePRL07}). However, the amplitudes of $I_{c}$ are different
because the models considered here and in Ref.\cite{BraudePRL07} are
different (a weak PE, a long JJ in our model and a strong PE and a short JJ
in Ref.\cite{BraudePRL07}). For $\alpha _{l}=\alpha _{r}$ the angle
dependence of the critical current $I_{c}$, Eq.(\ref{J4}), is the same as
obtained by Buzdin and Houzet \cite{HouzetPRB07} for a three magnetic layer
SF$_{l}$FF$_{r}$S Josephson junction. This model has been analysed recently
by Houzet and Birge in more detail \cite{Birge19} (see also \cite{Birge20}).
Similar angle dependence of the Josephson critical current was obtained,
mostly numerically, in Ref.\cite{HaltermanPRB14}.

\begin{figure}[tbp]
\caption{(Color online.) The dependence of the critical current on the field
$h$.}
\label{Crit.Current on h}
\end{figure}
In Fig.6 we show the normalized critical current $J_{c}=\mathcal{I}%
_{c}(\lambda )/\mathcal{I}_{c,n}$ as a function of $\lambda $. One can see
that the critical current reaches a maximum value at $\lambda \sim 1$ and
decreases to zero at large $\lambda $.

\subsection{Negative Josephson Current and Paramagnetic Response}

In this section we discuss the analogy between negative critical Josephson
current $I_{c}$ and a paramagnetic response of a superconducting system to
an external magnetic field. As we mentioned before, the negative $I_{c}$ may
arise in a magnetic S-F-S Josephson junctions and in multiterminal S-n-S
Josephson junctions with a nonequilibrium distribution function $\hat{n}%
(\epsilon )$. The negative $I_{c}$ in magnetic JJs has been predicted in
Refs.\cite{Bulaev77,BuzdinKup90} and observed in Refs.\cite%
{RyazanovPRL01,RyazanovPRL06}. In a recent paper \cite{LinderBelzigPRL20},
the possibility of a paramagnetic response of S-n bilayer with a
nonequilibrium distribution function was analyzed . Here we point out the
close analogy between negative $I_{c}$ and paramagnetic response. We show
that the response of a JJ with negative $I_{c}$ to external fields (ac
electric or magnetic) is paramagnetic regardless of the mechanism of
negative critical current. Indeed, it is well known that at low
temperatures, a JJ in an electric circuit plays a role of an inductance $%
\mathcal{L}$. For small variation $\delta \varphi =\varphi -\varphi _{0}$
and $I_{J},$ Eq.(\ref{J1}) can be written

\begin{equation}
\partial I_{J}/\partial t\cong I_{c}\partial (2\delta \varphi )/\partial
t\cos \varphi _{0}=I_{c}\frac{2eV}{\hbar }\text{.}  \label{JN1}
\end{equation}%
As follows from this equation, $\mathcal{L}=I_{c}\hbar /(2e)\cos \varphi
_{0} $. Thus, at a fixed $\varphi _{0}$ the inductance $\mathcal{L}$ changes
sign if $I_{c}$ becomes negative. On the other hand, the London equation
yields

\begin{eqnarray}
\partial I_{J}/\partial t &=&-\Lambda \partial A/\partial t=c\Lambda E=
\label{JN2} \\
&=&c(\Lambda /l_{ch})V\text{.}
\end{eqnarray}%
where $l_{ch}$ is a characteristic length which is determined by a concrete
type of \ a system. The effective inductance is $\mathcal{L}%
=l_{ch}/(c\Lambda )$. The positive coefficient $\Lambda $ corresponds to a
diamagnetic response while negative $\Lambda $ describes a paramagnetic
response. The negative inductance $\mathcal{L}$ means a paramagnetic
response of a JJ which has a negative $I_{c}$.

\begin{figure}[tbp]
\caption{(Color online.) The dependence of the integral $J$ on the ratio $%
R=P_{y}/P_{y}$.}
\label{J(R)}
\end{figure}

\section{Conclusions}

We have studied propagation of the LRSTC in a \textit{magnetic Andreev
interferometer}. The LRSTC is created by two thin ferromagnetic layers F$%
_{l,r}$ deposited on the superconductors S. For the propagation of the LRSTC
it does not matter whether the wires connecting the normal metal reservoirs
N or superconducting reservoirs S are made of normal (n) or magnetic (F)
metals. The magnetic layers F$_{l,r}$ have magnetisations $M_{l,r}$ which
are characterized by the angles $(\alpha )_{l,r}$ in the spin space. The
oscillating part of the dissipative current between the N reservoirs $%
I_{osc}=I_{V}\chi (\alpha ,\beta )\cos \varphi $ has the same angle
dependence as the Josephson current between the S reservoirs $%
I_{J}=I_{c0}\chi (\alpha )\sin \varphi $. However, the current $I_{osc}$
decreases with increasing temperature $T$ or the length $L_{x}$ much slower
than the critical current $I_{c0}$ (see Fig.6). In the first case the
decrease follows the power law behavior, while in the second case the
decrease is exponential: $I_{c0}\sim \exp (-2L_{x}/\xi (T))$. The critical
current $I_{c}=I_{c0}\chi (\alpha ,\beta )$ has different signs in
topological JJ's ($\alpha _{r}-\alpha _{l}=\pi $) and in non-topological
ones ($\alpha _{r}-\alpha _{l}=0$). At certain angles, the Josephson and
phase-dependent dissipative currents turn to zero, for example, for angles $%
\alpha _{r}-\alpha _{l}=(\pi /2)(2n+1)$ and $\beta _{r,l}=\pi /2$. Note that
we assumed that the proximity effect is weak. This is true if the parameters
$\kappa _{b,B}/\kappa _{\omega }\ll 1$. However the obtained results remain
qualitatively valid if this ratio is of the order of 1.

In the language of Ref.\cite{MoorVE15}, the obtained current-phase
dependence, $I_{J}=I_{c}\sin \varphi $, corresponds to a nematic case
contrary to a ferromagnetic case, $I_{J}=I_{c}\sin (\varphi +\psi )$, that
is, the Josephson current is equal to zero for the phase difference $\varphi
=0$. Therefore, it is of interest experimentally to investigate the angle
and phase dependence of the currents $I_{J}$ and $I_{osc}$. The obtained
results for the Josephson current $I_{J}$ are valid not only for the JJ
shown in Fig.1, but also for a planar geometry used in Ref.\cite%
{BirgePRL10,BirgePRL16}. Note also that the measurements of the $I_{osc}$ in
Andreev interferometers provides an additional possibility to study the
propagation of the LRSTC in magnetic superconducting structures.

\section{Acknowledgements}

The author thanks Ilya M. Eremin for careful reading of the manuscript and
acknowledge support from the Deutsche Forschungsgemeinschaft Priority
Program SPP2137, Skyrmionics, under Grant No. ER 463/10.

\section{Appendix}

\bigskip The formulas for the amplitudes $\mathcal{I}_{reg}$, $\mathcal{I}%
_{an}$ can be readily obtained from Eqs.(\ref{Anom}-\ref{Reg2}). We find
\begin{widetext}
\begin{eqnarray}
\mathcal{I}_{reg}(v) &=&\frac{\pi }{4}\text{Im}\sum_{n\geqslant 0}\{\frac{%
c_{reg}(\zeta _{n})\exp (-2\theta _{x}(\zeta _{n}))}{(\zeta _{n}+2iv)\cosh
^{2}\theta _{y}(\zeta _{n})}[\frac{\sinh (2\theta _{y}(\zeta _{n}))}{2\theta
_{y}(\zeta _{n})}-1]\}  \label{A1} \\
\mathcal{I}_{an}(v) &=&-\frac{1}{8}\frac{\sinh (2v)}{P_{y}}\int_{0}^{\infty }%
\frac{c_{an}(\zeta )d\zeta }{\zeta ^{3/2}}\frac{\text{Im}[(1-i)\tanh
(P_{y}(1+i)\sqrt{\zeta })]}{\cosh (\zeta +v)\cosh (\zeta -v)}\}\exp (-2P_{x}%
\sqrt{\zeta })\text{,}  \label{A1a} \\
\end{eqnarray}
\end{widetext} where $\theta _{x,y}(\zeta _{n})=P_{x,y}\sqrt{\zeta _{n}+2iv}
$. The functions $j_{reg,an}$ are given by equations
\begin{widetext}
\begin{eqnarray}
c_{reg}(\zeta _{n}) &=&-\frac{\lambda _{1,2}^{2}}{[\lambda _{1,2}^{2}+\sqrt{%
\zeta _{n}+2iv}]^{2}}(F_{S}^{R})^{2},  \label{A2} \\
c_{an}(\zeta ) &=&\frac{\lambda _{1,2}^{2}}{(\lambda _{1,2}^{2}+\sqrt{\zeta
})^{2}+\zeta }F_{S}^{R}F_{S}^{A}. \label{A2a}
\end{eqnarray}
\end{widetext}and the constants $\lambda _{1,2}$ are equal to: $\lambda _{1}%
\mathcal{=}\kappa _{l}/\sqrt{\kappa _{F}\kappa _{T}}$, $\lambda _{2}\mathcal{%
=}\kappa _{l}/\kappa _{T}$.

The constants $p_{i}$ and $p_{J1,2}$are defined as follows

\begin{eqnarray}
p_{1} &\mathcal{=}&\frac{1}{2}(\frac{\kappa _{B}\kappa _{b}}{\kappa _{T}^{2}}%
)^{2}(\frac{\kappa _{T}}{\kappa _{F}})\text{, }p_{2}\mathcal{=}\frac{1}{2}(%
\frac{\kappa _{B}\kappa _{b}}{\kappa _{T}^{2}})^{2}\text{,}  \label{Ap} \\
p_{J1} &\mathcal{=}&(\frac{\kappa _{b}}{\kappa _{T}})^{2}(\frac{\kappa _{T}}{%
\kappa _{F}})\text{, }p_{J2}\mathcal{=}(\frac{\kappa _{b}}{\kappa _{T}})^{2}%
\text{.}  \label{ApJ}
\end{eqnarray}

We also write the expression of the critical current $\tilde{I}%
_{c}=I_{c}(..) $ of a S - n - S Josephson junction with the same S/n
interface penetrability as in the considered structure. This quantity can
serve as a reference scale of the current

\begin{equation}
\delta \tilde{I}_{c,n}=p_{n}\mathcal{I}_{c,n}\sin \varphi \text{.}
\label{A3}
\end{equation}%
where $p_{n}=(\kappa _{b}/\kappa _{T})^{2}$ and the function $\mathcal{I}%
_{n} $ is

\begin{equation}
\mathcal{I}_{c,n}=2\pi \sum_{n\geqslant 0}\frac{\exp (-2P_{x}\sqrt{\zeta _{n}%
})}{\sqrt{\zeta _{n}}}\text{.}  \label{A3a}
\end{equation}%
with $\zeta _{n}=\pi (2n+1)$.

\bigskip


\begin{thebibliography}{999}
\bibitem{Footnote} The idea similar to that presented in Ref.\cite{BVEprl01}
was discussed a little later in Ref.\cite{Kadigr01}. The authors of Ref.\cite%
{BVEprl01} considered a diffusive case using the Usadel equation. The width
of the domain wall $d_{DW}$ in the ferromagnet was assumed to be larger than
the mean free path $l$. The authors of Ref. \cite{Kadigr01} considered the
case when $d_{DW}\lesssim l$. However they could not find the appropriate
correct solution for the Eilenberger equation. This solution for the case of
a narrow domain wall has been found in Ref.\cite{VolkovEfetovPRB08}.
\end{thebibliography}
\end{document}